\documentclass[%
 aip,
 sd,%
 amsmath,amssymb,
 reprint,%
 groupedaddress
]{revtex4-1}

\usepackage{graphicx}% Include figure files
\usepackage{dcolumn}% Align table columns on decimal point
\usepackage{mathtools}
\usepackage{bm}% bold math
\usepackage{color}
\usepackage{nameref}
\usepackage{hyperref}

\begin{document}

\preprint{AIP/123-QED}

\title{Equivalent Definitions of the Mie-Gr\"{u}neisen Form}

\author{Kirill A. Velizhanin}
 \email{kirill@lanl.gov}

\author{Joshua D. Coe}%
 \email{jcoe@lanl.gov}
\affiliation{Theoretical Division, Los Alamos National Laboratory, Los Alamos, NM, 87545}%

\date{\today}

\begin{abstract}
We define the Mie-Gr\"{u}neisen form in five different ways, then demonstrate their equivalence. 
\end{abstract}

\pacs{Valid PACS appear here}% PACS, the Physics and Astronomy
                             % Classification Scheme.
%Use showkeys class option if keyword
                              %display desired
\maketitle

\section{Introduction}

Equations of state (EOS) often are described as being of Mie-Gr\"{u}neisen (MG) form, or type. This is usually meant to convey that the pressure relative to that of some reference curve (subscripted `0') can be derived from the energy relative to that curve via
\begin{equation}
P(V, T) - P_0(V) = \frac{\Gamma(V)}{V} [E(V, T) - E_0(V)],
\label{eq:MG}
\end{equation}
where the Gr\"{u}neisen parameter $\Gamma$ is a function of volume only. MG EOS appear frequently in high-pressure and -temperature physics. They are particularly useful when applied to materials for which a single locus (such as the principal Hugoniot or isentrope) is well-characterized experimentally, with far fewer or lower-quality data elsewhere.  This is true of many materials with shock data recorded in the LASL\cite{marsh} and LLNL\cite{vanthiel} compendia, as well as most high explosive (HE) product mixtures.\cite{pointer}  Procedures for extracting the cold compression curve from shock data using Eq. \eqref{eq:MG} appeared very early in the history of shock physics,\cite{Benedek,rice} and fits of the Jones-Wilkins-Lee (JWL) form\cite{jwl} to cylinder expansion data can be made thermodynamically complete on the same basis.\cite{RalphJWL} 

While the MG form often is \emph{defined} through Eq. \eqref{eq:MG},\cite{eliezer} additional statements (discussed below) have been derived as consequences. Aside from the question as to whether \eqref{eq:MG} is actually overdetermined (\emph{i.e.}, redundant), many discussions implicitly suggest its primacy by the order in which alternative expressions are presented, as if they were merely its consequences. 

Our purpose here is to clarify the precise relationship between various formulae that have appeared in discussions of the MG form. Specifically, we will define the form in five different ways that do \emph{not} include \eqref{eq:MG}, then demonstrate their complete equivalence. Each pairwise relationship is one of mutual entailment, meaning that any single definition is sufficient to prove the others and that primacy cannot be assigned to any individual. Because no one definition is independent or unique, its emphasis in a particular context is purely a matter of convenience. Some of these results are well-known and have appeared elsewhere (at most depth in Ref.~\onlinecite{Menikoff-LA-UR-16-21706}), but we are aware of no previous attempts to so explicitly make these connections clear. 

The following section introduces the definitions, their equivalence is proven in Sec. \ref{sec:equiv}, and in Section \ref{sec:concl} we discuss some implications.

\section{Definitions}\label{sec:defs}

The following five definitions of the MG form are equivalent, meaning that any one serves as the necessary and sufficient condition for any other. Not every definition is expressed in a single equation, so definitions will be distinguished from equation numbers by a prepended `\#'.

\subsection*{Definition \#1}\label{sec:def1}

The first definition is a slight generalization of \eqref{eq:MG},
\begin{equation}
P(V, T) - P_0(V) = \frac{\Gamma}{V} [E(V, T) - E_0(V)],
\label{eq:MGf}
\end{equation}
where subscripts have the same meaning as before and $\Gamma$ is the Gr\"uneisen parameter defined by
\begin{equation}
\Gamma = V \left(\frac {\partial P}{\partial E} \right)_V = -\frac{V}{T}\left( \frac{\partial T}{\partial V} \right)_S.
\label{eq:G}
\end{equation}
Unlike \eqref{eq:MG}, \#1 imposes no restriction on the state-dependence of $\Gamma$ (\emph{viz.}, that it is a function only of $V$).

\subsection*{Definition \#2}\label{sec:def2}

In general, the Gr\"uneisen parameter as defined by Eq.~\eqref{eq:G} is an arbitrary function of the thermodynamic state. In the MG form, it is a function of volume only,
\begin{equation}
\Gamma=\Gamma(V).
\label{eq:G_V}
\end{equation}

\subsection*{Definition \#3}\label{sec:def3}

The entropy, 
\begin{equation}
S=f(x),
\label{eq:Sx}
\end{equation}
is a function only of the scaled temperature, $x=T/\Theta(V)$. The scaling factor $\Theta(V)$ is usually a characteristic temperature such as that of Debye or Einstein\cite{mcquarrie} and is, at most, a function of $V$.

\subsection*{Definition \#4}\label{sec:def4}
The heat capacity at constant volume,
\begin{equation}
C_V=\left(\frac{\partial E}{\partial T}\right)_V=T\left(\frac{\partial S}{\partial T}\right)_V,
\label{eq:CV}
\end{equation}
is a function only of the entropy,
\begin{equation}
C_V=C_V(S).
\end{equation}

\subsection*{Definition \#5}\label{sec:def5}

The temperature- and volume-dependencies of the Helmholtz free energy can be decomposed as 
\begin{equation}
A(V,T) = \phi(V) + \Theta(V) f(x),
\label{eq:helm}
\end{equation}
where $f(x)$ and $\phi(V)$ are single-argument functions that are otherwise arbitrary, and the same comments apply to $x=T/\Theta(V)$ as in \eqref{eq:Sx}. The first term on the right hand side is an internal energy as well as a free energy.

\section{Equivalence}\label{sec:equiv}

\begin{figure}[h]
\includegraphics[width=9cm]{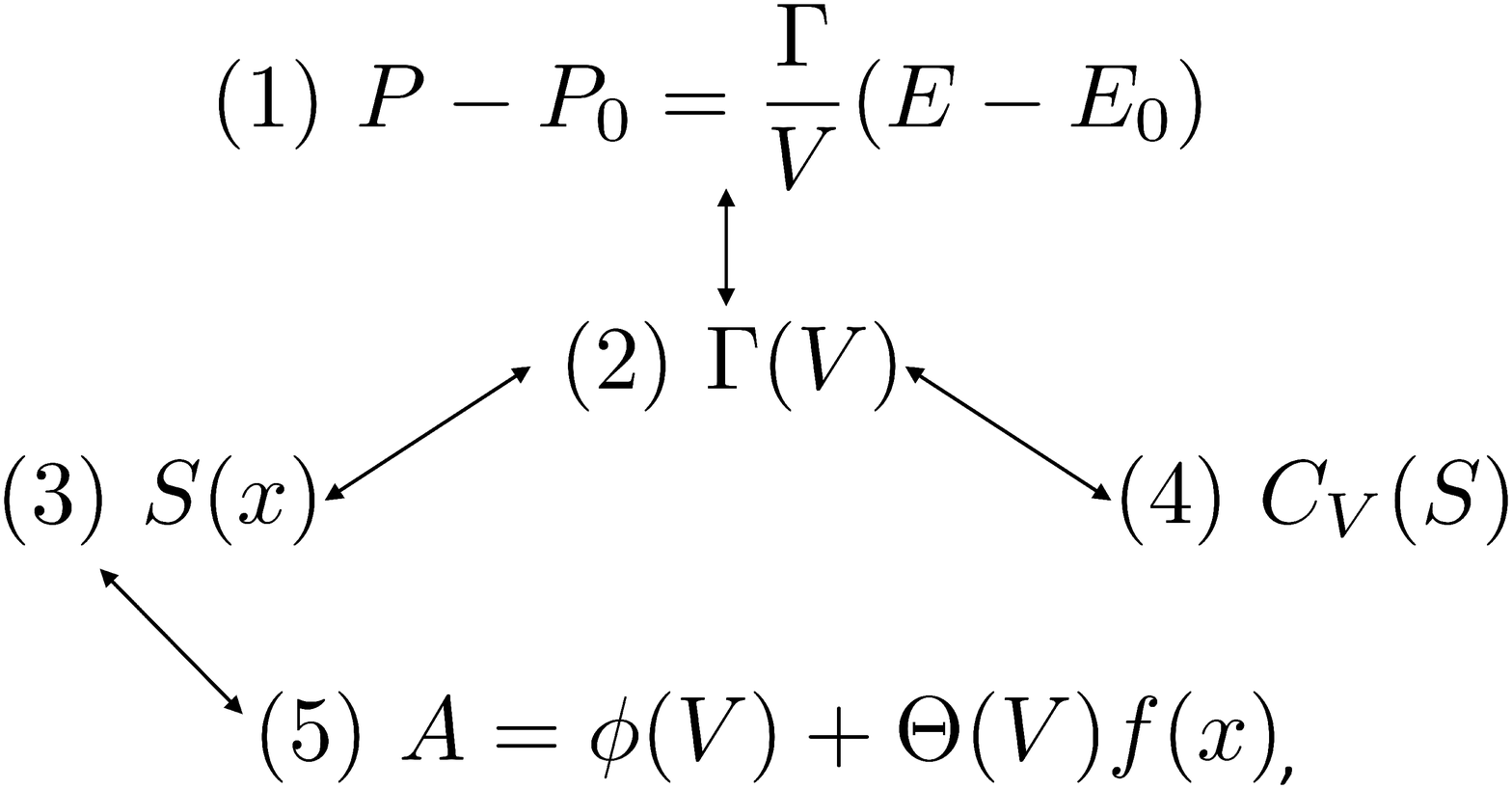}
\caption{\label{fig:diagram} Five definitions of the Mie-Gr\"{u}neisen form, with arrows indicating demonstrations of equivalence provided in the text. $\Gamma$ satisfies \eqref{eq:G}, and $x=T/\Theta(V)$.}
\end{figure}

Each of the following subsections demonstrates equivalence of a pair of definitions drawn from the five given above. Specifically, we consider four different pairs indicated by the arrows in Figure \ref{fig:diagram}. This diagram is complete in the sense that any definition follows from any other, either directly (\emph{e.g.}, \#2$\rightarrow$\#4) or via one additional step (\emph{e.g.}, \#3$\rightarrow$\#2$\rightarrow$\#1). More arrows could readily be drawn,  (\emph{e.g.}, \#3$\rightarrow$\#1), but we have opted for a sufficient set based on what seemed to be the simplest proofs. We also wish to emphasize that the topology of Figure \ref{fig:diagram} should not be misconstrued to assign any \emph{conceptual} significance to \#2, but merely reflects the convenience of its use.

\subsection{\#1 $\xleftrightarrow{}$  \#2}

Differentiation of Eq. \eqref{eq:MGf} with respect to $T$ at constant $V$ produces
\begin{equation}
\left(\frac{\partial P}{\partial T}\right)_{V}=\left(\frac{\partial\Gamma}{\partial T}\right)_{V}\frac{\left[E(T,V)-E_{0}(V)\right]}{V}+\frac{\Gamma}{V}C_V.
\label{eq:PTV} 
\end{equation}
The left-hand-side (lhs) of \eqref{eq:PTV} can also be evaluated using the chain rule in combination with Eqs. \eqref{eq:G} and \eqref{eq:CV},
\begin{equation}
  \left(\frac{\partial P}{\partial T}\right)_{V}=\left(\frac{\partial P}{\partial E}\right)_{V}\left(\frac{\partial E}{\partial T}\right)_{V} =\frac{\Gamma}{V}C_V.
\label{eq:PTV_chain}
\end{equation}
Equality of Eqs. \eqref{eq:PTV} and \eqref{eq:PTV_chain} requires that $\left(\frac{\partial \Gamma}{\partial T}\right)_V=0$, establishing that \#2 follows from \#1.

Conversely, integrating Eq. \eqref{eq:PTV_chain} with respect to $T$ at constant $V$ under the assumption that $\Gamma/V$ is a function of $V$ only (definition \#2) produces  
\begin{equation}
P=\frac{\Gamma}{V}E+C(V).
\end{equation}
Taking $C(V)=P_0(V)-\frac{\Gamma(V)}{V}E_0(V)$ for some reference curve $\{P_0(V),E_0(V)\}$ produces Eq. \eqref{eq:MGf}. Definition \#2 thereby entails \#1.

\subsection{\#2 $\xleftrightarrow{}$  \#3}

We first demonstrate \#3$\rightarrow$\#2. The Gr\"uneisen parameter can also be written in terms of observables as\cite{Menikoff-1989-75}
\begin{equation}
\Gamma = \frac{V \alpha B_T}{C_V},
\label{eq:gamprop}
\end{equation}
where $\alpha$ is the volumetric expansion coefficient, 
\begin{equation}
\alpha = \frac{1}{V} \left( \frac{\partial V}{\partial T} \right)_P,
\end{equation}
and $B_T$ is the isothermal bulk modulus,
\begin{equation}
B_T = -V \left( \frac{\partial P}{\partial V} \right)_T.
\end{equation}
Based on definition \#3, the denominator of \eqref{eq:gamprop} is
\begin{equation}
 C_V=T\left(\frac{\partial S}{\partial T}\right)_V=x \frac{dS}{dx},
 \label{eq:cv}
\end{equation}
and the numerator is
\begin{align}
 V \alpha B_T &= V \left( \frac{\partial P}{\partial T}  \right)_V \nonumber = V \left( \frac{\partial S}{\partial V} \right)_T \nonumber \\
 &= V \frac{dS}{dx}\left( \frac{\partial x}{\partial V} \right)_T = -\frac{VT}{\Theta^2} \frac{dS}{dx} \frac{d\Theta}{dV} \nonumber \\
 &= - x \frac{dS}{dx} \frac{d \ln \Theta}{d \ln V}.
 \label{eq:dpdt}
\end{align}
The first of the equalities in \eqref{eq:dpdt} follows from the cyclic rule for partial differentiation\cite{Menikoff-1989-75}
\begin{equation}
 \left( \frac{\partial x}{\partial y} \right)_z \left( \frac{\partial z}{\partial x} \right)_y \left( \frac{\partial z}{\partial x} \right)_y = -1,
\end{equation}
and the second from a Maxwell relation in the Helmholtz representation.\cite{callen} Substitution of \eqref{eq:cv} and \eqref{eq:dpdt} into \eqref{eq:gamprop} gives
\begin{equation}
 \Gamma = - \frac{d\ln \Theta}{d \ln V},
 \label{eq:gam_quasi}
\end{equation}
a function only of $V$. This argument is very similar to that of Wallace,\cite{wallacesxn} and is sufficient to prove that \#2 follows from \#3.

The converse, \#2$\rightarrow$\#3,  is slightly more detailed.
Again using the cyclic rule, the second definition of $\Gamma$ in \eqref{eq:G} can be rewritten as
\begin{equation}
\Gamma(V) = -\frac{V}{T} \left(\frac{\partial T}{\partial V} \right)_S = \frac{V}{T} \left(\frac{\partial T}{\partial S} \right)_V \left(\frac{\partial S}{\partial V} \right)_T,
\end{equation}
or as a partial differential equation for the entropy, 
\begin{equation}
  \frac{V}{\Gamma(V)} \left(\frac{\partial S}{\partial V}\right)_T - T \left(\frac{\partial S}{\partial T}\right)_V = 0.
  \label{eq:pde}
\end{equation}
Equation \eqref{eq:pde} can be solved by the method of characteristics, where the characteristics satisfy the following system of ordinary differential equations parameterized by $t$:
\begin{equation}
  \frac{dV}{dt} = \frac{V}{\Gamma(V)}
\end{equation}
\begin{equation}
  \frac{dT}{dt} = -T
\end{equation}
\begin{equation}
  \frac{dS}{dt} = 0.
\end{equation}
The last of these demonstrates that the characteristic curves are isentropic, whereas the first two can be combined to eliminate $t$ 
\begin{equation}
\frac{dT}{T} + \frac{\Gamma(V)}{V} dV = 0,
\end{equation}
and then straightforwardly integrated via separation of variables, yielding
\begin{equation}
 T e^{\int_{V_0}^{V} \frac{\Gamma(V')}{V'} dV'} = C.
\end{equation}
The solution to \eqref{eq:pde} is the union of all such curves (conveniently labeled by the integration constant $C$) and entropy can thus be expressed as
\begin{equation}
 S = f(C)=f \left( T e^{\int_{V_0}^{V} \frac{\Gamma(V')}{V'} dV'} \right)
\end{equation}
for arbitrary function $f$. This expression is simply Eq. \eqref{eq:Sx} with $\Theta(V)=e^{-\int_{V_0}^{V} \frac{\Gamma(V')}{V'} dV'}$, thereby proving \#2 $\rightarrow$ \#3.

\subsection{\#2 $\xleftrightarrow{}$ \#4}

The equality of these statements was shown originally by Davis\cite{davis} and discussed in more detail by Menikoff.\cite{Menikoff-LA-UR-16-21706} It is based on the invariance of mixed third derivatives of $E(S,V)$ to the order in which they are taken (\emph{i.e.}, a higher-order analogue of the Maxwell relations). In this case,
\begin{equation}
\frac{\partial}{\partial V} \frac{\partial^2 E}{\partial S^2} = \frac{\partial^2}{\partial S^2} \frac{\partial E}{\partial V},
\end{equation}
which can be reexpressed as
\begin{equation}
-\frac{\Gamma}{V}\frac{T}{C_V} \left( \frac{\partial \Gamma}{\partial S} \right)_V = \frac{V}{C_V^2} \left( \frac{\partial C_V}{\partial V} \right)_S.
\end{equation}
The lhs vanishes if $\Gamma$ is a function only of $V$, requiring that $C_V$ be a function only of $S$ and vice-versa.

\subsection{\#3 $\xleftrightarrow{}$ \#5}

That \#5$\rightarrow$\#3 follows from simple differentiation, $\left(\frac{\partial A}{\partial T}\right)_V=-S$. Conversely, integration of $S=f(T/\Theta(V))$ with respect to $T$ along some path of constant volume gives
\begin{equation}
A(T,V)-A_0(V)=\theta(V)g(T/\Theta(V)),
\end{equation}
where $\frac{dg(x)}{dx}=-f(x)$. The function $g(x)$ is known up to an integration constant. Convenient choices for this constant include $g(0)=0$ and $\theta(V)g(0)=E_{ZP}$, where $E_{ZP}$ is the vibrational zero-point energy (zpe). The first results in $\left.A(V,T)\right|_{T=0}=A_0(V)$ equivalent to the zero-temperature isotherm, the second in its being the cold internal energy without zpe. The latter is often calculated using density functional theory,\cite{martin} and is sometimes referred to as the cold curve.\cite{peterson}

\section{Comments}\label{sec:concl}

\begin{itemize}
\item Definitions \#1 and \#2 often are combined as the single statement \eqref{eq:MG}, when in fact this is redundant: \#1$\leftrightarrow$\#2 and so either alone is sufficient to fully define the MG form. 

\item It is sometimes convenient to generalize Eq.~\eqref{eq:MG} to $\Gamma$ as a function of variables in addition to volume, such as internal energy or temperature.\cite{fortov} This poses no difficulty so long as it is understood that the new $\Gamma(V,E)$ or $\Gamma(V,T)$ is \emph{not} that defined by Eq. \eqref{eq:G}.  For $\Gamma$ to satisfy both \eqref{eq:MGf} and \eqref{eq:G}, it must also satisfy \eqref{eq:G_V} because \#1$\rightarrow$\#2. See, for example, Eqs. (1a-b) and (2a-b)  of Ref.~\onlinecite{fumi}.

\item If the entropy is invertible, then \#4 can be recast in the (possibly more intuitive) form $C_V = C_V(x)$. Generally speaking, this should be true except at phase boundaries.

\item We have restricted our discussion to various expressions of the MG form at the level of classical thermodynamics, without having considered the microscopic implications of their validity. The literature on this topic is vast, but the work most closely related to ours is Ref. \onlinecite{fumi}. Such considerations typically are framed in terms of what is required in order for \#2 to hold.

\item For systems poorly-described by a single characteristic temperature (\emph{e.g.}, polymers\cite{wunderlich} or other molecular solids such as high explosives\cite{cawkwell}), it is often convenient to express the free energy as a sum of non-interacting components,
\begin{equation}
A(V,T)=\sum_i A_i(V,T),
\end{equation}
where each component free energy can be decomposed in the manner of \eqref{eq:helm}. Fig. \ref{fig:diagram} and all proofs still apply separately to each component $i$.  For example, the partial pressures are
\begin{equation}
P_i-P_{i,0}=\frac{\Gamma_i}{V}[E_i-E_{i,0}],
\end{equation}
where $\Gamma_i=-V\frac{d\Theta_i(V)}{dV}$ and the total pressure $P=\sum_i P_i$. Note, however, that the thermodynamically consistent definition of $\Gamma$ is not the sum of $\Gamma_i$. Rather, the $\Gamma$ consistent with \eqref{eq:G} and \eqref{eq:gamprop} is given by
\begin{equation}
\Gamma=\frac{\sum_i C_{V,i}\Gamma_i}{\sum_i C_{V,i}},
\end{equation}
where
\begin{equation}
C_V = \sum_i C_V,i.
\end{equation}
Multiple characteristic temperature models can therefore be of MG form only if all characteristic temperatures have the same $\Gamma$.

\end{itemize}

\bibliography{library}

\end{document}